# The Usefulness of Type Ia Supernovae for Cosmology— A Personal Review

**Kevin Krisciunas**
*George P. and Cynthia Woods Mitchell Institute for Fundamental Physics and Astronomy, Department of Physics and Astronomy, Texas A & M University, 4242 TAMU, College Station, TX 77843; krisciunas@physics.tamu.edu*

**Abstract**   We review some results of the past twelve years derived from optical and infrared photometry of Type Ia supernovae. A combination of optical and infrared photometry allows us to determine accurately the extinction along the line of sight. The resulting distance measurements are much more accurate than can be obtained from optical data alone. Type Ia supernovae are very nearly standard candles in the near-infrared. Accurate supernova distances, coupled with other observational data available at present, allow us to determine the matter density in the universe and lead to evidence for the existence of Dark Energy. We can now address some questions on the grandest scale such as, "What is the ultimate Fate of the universe?"

## 1. Introduction

The basic supernova (SN) classification scheme stipulates that supernovae (SNe) with hydrogen emission in their spectra are called Type II SNe, and those without hydrogen are called Type I SNe (Minkowski 1941). For more than twenty years we have delineated subclasses. Type Ia SNe show absorption due to singly ionized silicon. The key line has a rest wavelength of 6355Å (Filippenko 1997). As it is observed at roughly 6150Å at the time of maximum light, this signifies that a typical outflow velocity at maximum light is about 10,000 km s$^{-1}$.

It is generally believed that a Type Ia SN is caused by the explosion of a carbon-oxygen white dwarf (WD) star (in a close binary system) which has approached the Chandrasekhar limit of $1.4\,M_\odot$. Two possible scenarios are envisioned. Either the WD gains mass from a nearby donor star (a main sequence star or a giant), or the explosion results from the merging of two white dwarfs.

In Figure 1 we show a schematic diagram of the light path from a SN to a telescope on the Earth. At maximum light (roughly 19 days after the explosion) the size of the expanding fireball is ~100 AU. Light is possibly scattered by circumstellar material near the SN (Wang 2005; Goobar 2008). The light may be dimmed by dust in the host galaxy. The light waves are shifted towards longer wavelengths owing to the expansion of the universe, given the redshift of the host galaxy. The light is dimmed by dust in the Milky Way galaxy. Finally, the light passes through the Earth's atmosphere, is reflected and/or transmitted by the optical elements of the telescope, is spread out into a spectrum or passes through one of several broad band filters, and hits the light detector in our spectrograph or CCD camera, the quantum efficiency of which is a function of wavelength. Thus, the SN light is affected by many factors, and if we are to understand SNe well, we must standardize our spectra and broad band photometry.

Phillips (2012) has written an excellent review of the near-infrared (IR) properties of Type Ia SNe. We can hardly improve on his article. Here we shall emphasize our contributions over the past dozen or so years. One topic we shall not discuss here is the morphology of Type Ia SN lightcurves. We refer the reader to Hamuy *et al.* (1996), Riess *et al.* (1996), Kasen (2006), Jha *et al.* (2007), and the relationship between the *B*-band decline rate parameter and the strength of the *I*-band secondary maximum shown in Figure 17 of Krisciunas *et al.* (2001). Mandel *et al.* (2009) and Mandel *et al.* (2011) give sophisticated analysis of optical and near-IR light curves.

## 2. Standardizable candles and standard candles

In order to determine the distance to an astronomical object we make use of the standard relationship between absolute magnitude (M), apparent magnitude (m), and distance (d) in parsecs:

$$M = m + 5 - 5 \log (d), \qquad (1)$$

where the apparent magnitude is corrected for any dust extinction along the line of sight. A century ago Henrietta Leavitt discovered that Cepheids with longer periods are brighter than those with shorter periods (Leavitt and Pickering 1912). This is the famous period-luminosity relation. In short, its significance is that while Cepheids do not all have the same intrinsic brightness, those of a given stellar population type show a specific linear relation between the logarithm of the period in days and the mean absolute magnitude. Thus, they are standardizable candles. The brightest Cepheids are more than 10,000 times more luminous than the Sun. By contrast, RR Lyrae stars of a particular metallicity pulsating in the fundamental radial mode have mean visual absolute magnitudes of about +0.7 (Layden *et al.* 1996). They are standard candles—all exhibit the same luminosity, like 100-watt light bulbs only much brighter. RR Lyr stars are roughly forty times more luminous than the Sun. Type Ia SNe are very useful for extragalactic astronomy because at maximum light they are four billion times more luminous than the Sun; they can be detected halfway across the observable universe with a 4-meter class telescope.

Baade (1938) first suggested that Type I SNe may be useful for measuring accurate cosmic distances. (The subclass of Type Ia SNe was first identified by Elias *et al.* (1985).) Kowal (1968) made the first test of the usefulness of these objects as distance indicators. Phillips (1993) discovered the "decline rate relation" and established the usefulness of the parameter $\Delta m_{15}(B)$, which is the number of *B*-band magnitudes that a Type Ia SN declines in the first 15 days after *B*-band maximum; a typical value is 1.1 magnitudes, but the range is from 0.68 (Krisciunas *et al.* 2011) to 1.93 or so (Garnavich *et al.* 2004). As with Cepheids, the absolute magnitude is related to another observable. The most slowly declining Type Ia SNe are about 2.5 magnitudes more luminous in the *B*-band at maximum light than the fast decliners. The steepness of the decline rate relation becomes shallower as we proceed to longer wavelength bands (see Figure 2). Thus, at optical wavelengths, Type Ia SNe are standardizable candles.

Meikle (2000) presented a very useful compilation of IR photometry of Type Ia SNe published up to that time. Of particular note are the two papers of Elias *et al.* (1981, 1985); the latter paper presented the first IR Hubble diagram of Type Ia SNe, using the *H*-band magnitudes at 20 days after maximum light. Meikle (2000) also showed, on the basis of a small number of objects, that Type Ia SNe at ~14 days after maximum light may be standard candles. Krisciunas *et al.* (2003) also showed that Type Ia SNe might be standard candles in the near-IR. We based this on a sample of Type Ia SNe only slightly bigger than that of Peter Meikle, on the basis of their *H*-band (1.65 micron) magnitudes 10 days after *B*-band maximum.

One might ask why there are so many fewer near-IR data compared to optical photometry. Near-IR chips have fewer pixels than optical CCD cameras. A small field of view makes it more difficult to make mosaic images, as there would be fewer field stars of appropriate brightness near any SN. This is particularly important for images taken on non-photometric nights. Also, there are few telescopes systematically used for near-IR observations of SNe. Amongst them are the CTIO 1.3-meter telescope, the Las Campanas 1.0-meter and 2.5-meter telescopes, a 1.3-meter telescope at Mt. Hopkins, Arizona, and the Liverpool JMU 2-meter telescope at La Palma.

**3. Uniformity of color curves for determining extinction**

I began observing Type Ia SNe early in 1999 using the Apache Point Observatory 3.5-meter telescope. My collaborators were Gene Magnier, Chris Stubbs, and Alan Diercks of the University of Washington. (Stubbs and Diercks were members of the High-Z Supernova Team, whose highly cited paper (Riess *et al.* 1998) garnered a Nobel Prize in physics for Brian Schmidt and Adam Riess.) We published a paper (Krisciunas *et al.* 2000) which had two primary results, one dealing with color curves, the other dealing with unusual dust properties.

We found that Type Ia SNe whose decline rate parameter $\Delta m_{15}(B)$ was in the middle range delineated uniform optical vs. infrared color curves. This follows up a suggestion by Elias *et al.* (1985) that *V–K* colors of Type Ia SNe may be quite uniform. An example of *V–K* color curves of Type Ia SNe is shown in Figure 3. Using our data of SN 1999cp and data of SNe 1972E, 1980N, 1983R, 1981B, and 1981D (see Elias *et al.* 1981, 1985, and data and references given by Meikle 2000) we constructed a "zero reddening" locus. Optical and IR data for SN 1998bu (Suntzeff *et al.* 1999, Jha *et al.* 1999, Hernandez *et al.* 2000) show the same basic color curve shape, but reddened in the *V–K* color index by nearly 1 magnitude. (The *UBVRIYJHK* bands are at 0.36, 0.44, 0.55, 0.65, 0.80, 1.03, 1.25, 1.65, and 2.2 microns, respectively.) Our observations of SN 1999cl (Krisciunas *et al.* 2000, 2006) show that this object was reddened even more.

We can parameterize dust reddening as follows. Say we somehow know the unreddened *B–V* color of a star or SN. The difference of the observed color and the unreddened color is the color excess E(*B–V*). The *V*-band extinction is related to this color excess as follows:

$$A_V = R_V\, E(B\!-\!V). \qquad (2)$$

The standard value of $R_V = 3.1$ is for Milky Way dust (Cardelli *et al.* 1989), but this value can range from 1.5 to 5 depending on the line of sight in our Galaxy.

Extinction by interstellar dust is diminished at longer wavelengths. An analog of Equation 2, but using the *V*- and *K*-bands, is as follows:

$$A_V = \alpha\, E(V\!-\!K). \qquad (3)$$

where parameter $\alpha$ is in the range 1.08 to 1.14.

Even for a wide range of dust grain size and composition the scaling parameter in Equation 3 has a very small range. Thus, if one can obtain a *V–K* color excess, increasing that by ten percent gives us the *V*-band extinction. How much extinction one expects for that SN in other bands can be obtained using the coefficients calculated by Jose Prieto and given in Table 8 of Krisciunas *et al.* (2006).

Krisciunas *et al.* (2004b) present coefficients to generate the *V–H* and *V–K* color curves of the mid-range decliners, and the *V–J*, *V–H*, and *V–K* color curves of the slowly declining Type Ia SNe. This paper also gives the coefficients to generate *JHK* light curve templates valid from 12 days before the time of *B*-band maximum until 10 days after T($B_{max}$). While the fast-declining Type Ia SNe are considerably redder than more slowly declining objects of this type around maximum light, from 30 to 80 days after T($B_{max}$) Type Ia SNe of all decline rates show a certain uniformity in the *V–H* and *V–K* color indices (Krisciunas *et al.* 2009b). (In optical bands researchers make use of the "Lira Law." It relates to the uniformity of the unreddened *B–V* colors of Type Ia SNe from 32 to 92 days after T($B_{max}$) (Lira 1995, Phillips *et al.* 1999).) The uniformity of *V–H* and *V–K* colors of Type Ia SNe is backed up by modeling calculations by Peter Hoeflich and shown in Figure 12 of Krisciunas *et al.* (2003).

Our observations of SN 1999cl (Krisciunas *et al.* 2000, 2006) indicated that $R_V \approx 1.55 \pm 0.08$ for the host galaxy dust. Since then a small number of highly reddened Type Ia SNe have been observed, amongst them SN 2002cv (Elias-Rosa *et al.* 2008), 2003cg (Elias-Rosa *et al.* 2006), and SN 2006X (Wang *et al.* 2008). As Wang (2005) and Goobar (2008) point out, the light of a highly dimmed and reddened object can suffer from extinction and scattering. What the balance is of these processes is not understood at this time. With intrinsic color variations, which are expected for any group of cosmic objects, we thus have three sources of color effects.

Suffice it to say that adopting standard Galactic reddening of $R_V = 3.1$ for all Type Ia SNe is just wrong. The host of SN 1999cl, for example, is M88 in the Virgo cluster. Adopting $R_V = 3.1$ for the dust that affected SN 1999cl gives us a distance value that is halfway to the center of the Virgo cluster. Thus, either M88 is by chance in the same direction as the Virgo cluster, but not in it, or we need to adopt the value of $R_V$ derived from a combination of optical and IR data.

**4. Clones**

Krisciunas *et al.* (2007) found that, for all intents and purposes, SN 2004S was a clone of the well studied object SN 2001el (Krisciunas *et al.* 2003). Since the former is essentially unreddened in its host, we can correct both objects for a small amount of Milky Way dust extinction and determine that the host galaxy dust of SN 2001el was characterized by $R_V = 2.15 \pm 0.24$ and that SN 2001el suffered $0.47 \pm 0.03$ magnitude more *V*-band extinction than SN 2004S (see Figure 4). This result exploits the advantage of using a combination of optical and infrared photometry of Type Ia SNe. Previously, if we were limited to using only *B*- and *V*-band photometry, the uncertainty in distance to a reddened SN might have been ±20 percent, but by using optical and IR data the uncertainty of the extinction corrections leads to uncertainties in distance that can be as small as the random errors of the photometry, a few percent. This is a considerable improvement!

Type Ia SNe are not the only exploding stars that show certain uniformities of their lightcurves and color curves. The Type II-P SNe 1999em and 2003hn were found to be near-clones of each other. This allowed us to use the optical and IR photometry to calculate that SN 2003hn was dimmed by $0.25 \pm 0.03$ magnitude more in the *V*-band than SN 1999em (Krisciunas *et al.* 2009a).

**5. The first Hubble diagram of Type Ia SNe at maximum light in the near-IR**

Our light curve templates for the near-IR *JHK* bands (Krisciunas *et al.* 2004b, Table 12) allowed us to derive the maximum magnitudes of Type Ia SNe as long as there were some observations between twelve days prior to $T(B_{max})$ and ten days afterward. We used our *V* minus near-IR color curves to correct these apparent magnitudes at maximum light for extinction along the line of sight. This led to the first Hubble Diagram of Type Ia SNe at maximum light in the IR (Krisciunas *et al.* 2004a) (see Figure 5). The fact that the data fit the three straight lines like beads on a string means two things. One is not a surprise, that light intensity decreases as the square of the distance. But the other will be significant for all future surveys of Type Ia SNe, namely that they are better than standardizable candles in the IR. This sample of objects shows that they are standard candles. Of course, one wants a sample bigger than sixteen objects, but this was a good start.

The scatter in the near-IR Hubble diagrams obtained so far is about ± 0.15 magnitude, comparable to what one finds for *BVRI* Hubble diagrams. However, as we push out into the Hubble flow (that is, redshift $z > 0.01$) we can expect the near-IR Hubble diagrams to have tighter fits because of the minimal systematic errors in the extinction corrections and the diminished effect of the peculiar velocities.

## 6. Deviations from uniform standard candle nature

In Figure 5 if there were any points above the lines, that would indicate SNe intrinsically fainter than the rest, and points below the lines would indicate SNe intrinsically brighter. More information on the standard candle nature of Type Ia SNe can be gleaned from a plot of the absolute magnitudes at maximum brightness vs. some other parameter. Figure 6 is a plot from Krisciunas *et al.* (2011), but we have added three regression lines to subsets of the data. Figure 6 shows that Type Ia SNe are (nearly) perfect standard candles in the near-IR. SN 2009dc may have been a "super-Chandra" event, rather than a more standard Type Ia SN that produces roughly 0.5 $M_\odot$ of $^{56}$Ni. In our paper on SN 2003gs (Krisciunas *et al.* 2009b) we also used data of SN 1986G (Frogel *et al.* 1987) and four objects from the Carnegie Supernova Project (Contreras *et al.* 2010) to show that there is a bifurcation in the absolute magnitudes at peak brightness of the fast decliners. At the right hand side of Figure 6 the diamond shaped symbols correspond to objects that peaked in the near-IR after $T(B_{max})$. We have excluded these points and SN 2009dc from the regression lines shown in Figure 6. These lines have non-zero slopes only at the 1.3- to 2.2-σ levels of significance. (A 3-standard deviation result is usually the criterion for statistical significance. For a random sample of data this would occur only 0.5 percent of the time.) The data indicate that Type Ia SNe with $\Delta m_{15}(B) = 1.4$ are roughly 0.10–0.15 magnitude fainter in the near-IR than those with $\Delta m_{15}(B) = 0.8$. This is comparable to the uncertainties of the absolute magnitudes shown in Figure 6.

Folatelli *et al.* (2010, Figure 17) showed that there may be a non-zero slope to the *J*-band decline rate relation. More extensive data from the Carnegie

Supernova Project (Kattner *et al.* 2012) indicate non-zero slopes at the 2-σ level for the *YJH* bands.

Suffice it to say that Type Ia SNe at maximum brightness are excellent objects for determining extragalactic distances. They are nearly perfect standard candles in the near-IR. Excluding possible super-Chandra events and late-peaking subluminous objects, the slopes of the regression lines in Figure 6 are not statistically significantly different than zero.

**7. Hubble Diagrams and evidence for Dark Energy**

We do not have the space here to review the subject of high redshift SNe and the discovery of the acceleration of the universe. For a cosmology primer see the article "Fundamental cosmological parameters" (Krisciunas 1993, and references therein); also the discussion of "luminosity distances" in the Introduction to Krisciunas *et al.* (2005).

Two fundamental parameters used by observational cosmologists are the mean density of matter in the universe compared to the critical density, $\Omega_M$, and the Dark Energy density parameter, $\Omega_\Lambda$. If these two parameters sum to 1.00, then the geometry of the universe is flat. In Figure 7 we show the "distance modulus" (or $m-M$ from Equation 1) as a function of the logarithm of the redshift. Different models of the universe are shown. The "empty universe" has $\Omega_M = 0.0$, $\Omega_\Lambda = 0.0$. The "Einstein-de Sitter universe" (or critical density model) coasts to a stop after an infinite amount of time and has $\Omega_M = 1.0$. Prior to 1998 the expectation was that high redshift Type Ia SNe would show that $\Omega_M = 0.3$, $\Omega_\Lambda = 0.0$; this we call the "open model", as it would lead to the perpetual expansion of the universe even without a positive Cosmological Constant.

Figure 8 is a "differential Hubble diagram." We take the empty universe model from Figure 7 as a reference and plot the differences of the other loci with respect to the empty universe model. What Riess *et al.* (1998) and Perlmutter *et al.* (1999) found was that the high redshift SN data do not fall along the locus of the "open" model. Instead the SNe are "too faint" by about 0.19–0.25 magnitude from redshift 0.4–0.8. Possible explanations are: 1) that some kind of gray dust is dimming the light but not reddening it; 2) Type Ia SNe at this lookback time are inherently dimmer than nearby, more recent, Type Ia SNe; or 3) the SNe are further away than we would expect, given their redshifts, which is evidence for repulsive Dark Energy.

Riess *et al.* (2004) used the Hubble Space Telescope to find Type Ia SNe at even greater redshifts and found that the SNe at $z > 1.3$ were "too bright" compared to the empty universe model. This means that they looked far back enough in time to see the universe when it was small enough and dense enough that the gravitational attraction of matter on all other matter was stronger than any repulsive effect of a positive Cosmological Constant. At a redshift beyond 1.3 the universe is observed to be decelerating. The findings of Riess *et al.*

(2004) also proved that the faintness of Type Ia SNe at redshift ~0.5 was not due to some weird kind of gray dust.

We note that SN data alone do not give us enough leverage to determine the most accurate values of $\Omega_M$ and $\Omega_\Lambda$. Wood-Vasey *et al.* (2007) and others used SN data combined with information from "baryon acoustic oscillations" (Eisenstein *et al.* 2005). The flatness of the geometry of the universe is best demonstrated from the characteristic angular size of the warmer and cooler spots of the Cosmic Microwave Background (CMB) radiation, such as shown by the analysis of seven years of data from the Wilkinson Microwave Anisotropy Probe (WMAP) by Komatsu *et al.* (2011).

Once we know the matter and Dark Energy content of the universe, we can determine the expansion history of the universe (Figure 9). The universe was dense enough for the first seven billion years that the gravitational attraction of all the matter caused the expansion to be decelerated. After that the effect of repulsive Dark Energy has caused an acceleration of the expansion.

## 8. Future analysis required

Hicken *et al.* (2009) presented data for 185 nearby Type Ia SNe observed by astronomers from the Harvard-Smithsonian Center for Astrophysics. Only thirty-one of them have values of $\Delta m_{15}(B)$, maximum *U*-band magnitudes, and are at a redshift greater than $z = 0.01$ (which is regarded to be the beginning of the smooth Hubble flow). The *U*-band data show a scatter of ± 0.25 magnitude for a decline rate relation graph or a Hubble diagram, which is almost twice the scatter one sees in other photometric bands. Some of this extra scatter may be due to a viewing angle effect (Maeda *et al.* 2010). It is known that some Type Ia SNe are polarized, implying that the explosions are not spherically symmetric. Since the *U*-band light is dimmed more than the longer wavelength bands, this could increase the scatter of the derived *U*-band absolute magnitudes.

The other factor affecting the *U*-band data is the perennial challenge to correct the photometry for differences in the effective bandpasses used for all the observations from all telescopes for a given SN. Using lab data obtained by us, or provided by manufacturers, and spectra of the SN themselves, the method of spectroscopically-derived corrections (the so-called "S-corrections") allows us to resolve this problem, in principle. The method was originated by Stritzinger *et al.* (2002), who applied it to SN 1999ee, and by Krisciunas *et al.* (2003), who applied it to SN 2001el. Other papers written by us show photometric corrections to optical and IR photometry of many objects (Candia *et al.* 2003, Krisciunas *et al.* 2004b, 2004c, 2007, 2009b, 2011).

Our paper on SN 2003gs (Krisciunas *et al.* 2009b), however, did not include *U*-band S-corrections, and we chose at that time not to publish *U*-band photometry from one telescope because it was systematically 0.4 magnitude brighter at one month after maximum light compared to data from two other telescopes. We now have worked out corrections and can reconcile the otherwise discordant *U*-band data of SNe 2003gs and 2003hv obtained with three telescopes. This is a step in the right direction. Details will be published in a separate paper.

Why are the *U*-band data so important? Both the Sloan Digital Sky Survey supernova search (Kessler *et al.* 2009) and the CFHT Legacy Supernova Survey (Conley *et al.* 2011) discovered systematic errors in the distance moduli of high-redshift SNe when anchored with *U*-band photometry of nearby objects. Both projects decided to eliminate from the analysis data that originated in the restframe *U*-band. (For example, a SN at redshift 0.7 observed in the *R*-band gives us photons emitted in the *U*-band.) To utilize fully the SN surveys of the future such as the Dark Energy Survey and data from the Large Synoptic Survey Telescope we need to fix the old *U*-band photometry (which may be impractical or impossible), or we have to restrict ourselves to data obtained with a minimum number of telescopes and cameras, such as the Carnegie Supernova Project (Hamuy *et al.* 2006).

Until recently the effective filter profiles for S-corrections were obtained using laboratory data for the transmissions and reflectances of all the optics in a telescope and camera, then multiplying all these functions of wavelength together. This allowed us to reconcile previously discordant data obtained with cameras having significantly different filters. Stubbs *et al.* (2007) and Rheault *et al.* (2010) have shown the way to the future for SN calibration. They designed two systems to measure the effective filter profiles in situ. Stubbs *et al.* used a tunable laser and Rheault *et al.* use a "monochromator" to scan the transmission throughout the whole system from the ultraviolet through the near-IR.

**9. Conclusions**

Data obtained over the past decade have confirmed the suggestion of Elias *et al.* (1985) that optical minus infrared colors of Type Ia SNe are uniform within certain ranges of the *B*-band decline rate. A combination of optical and near-IR photometry allows us to determine the amount of extinction very well, and allows us to determine the reddening parameter $R_V$ (Krisciunas *et al.* 2007). The suggestion of Meikle (2000) and Krisciunas *et al.* (2003) that Type Ia SNe are nearly standard candles in the near-IR has been borne out by subsequent analysis of the absolute magnitudes at maximum light (Krisciunas *et al.* 2004a, Wood-Vasey *et al.* 2008, Kattner *et al.* 2012). The goal of future photometry of Type Ia SNe will be to obtain well calibrated data at ultraviolet, optical, and IR wavelengths for nearby SNe and also for SNe out to redshift $z = 2$. Some of these observations must be made from space, such as with the satellites Euclid, WFIRST, or the James Webb Space Telescope. Infrared data in particular, whether in the observer's frame or the restframe, will be very important for observational cosmology.

**10, Acknowledgements**

The author thanks Max Stritzinger and Mark Phillips for reading a previous draft of this paper, and for useful comments.

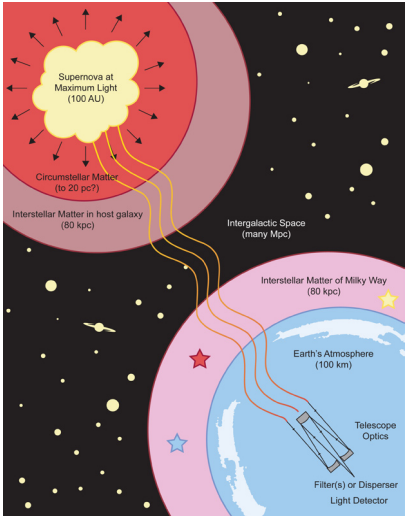

Figure 1. Schematic diagram of the light path of a Type Ia supernova to a telescope situated on the Earth. Figure by Elisabeth Button.

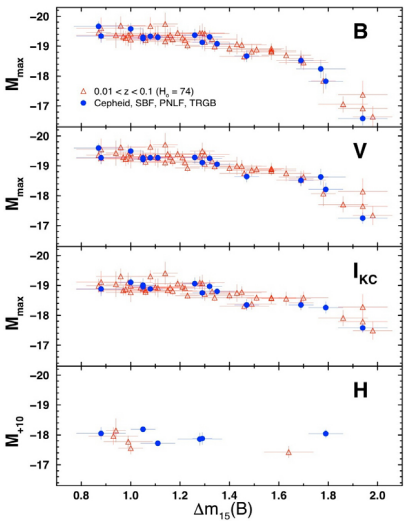

Figure 2. Decline rate relations of Type Ia SNe for the BVI and H-bands (Krisciunas et al. 2003). The x-axis parameter is the number of B-band magnitudes that a Type Ia SNe declines in the first 15 days after the time of B-band maximum. For BVI the absolute magnitudes are the maximum brightness values. For the near-IR H-band the absolute magnitudes are measured at 10 days after T(Bmax). Figure by Mark Phillips.

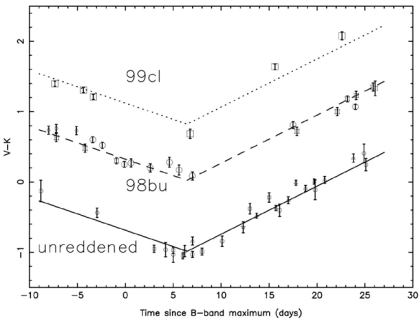

Figure 3. *V–K* colors of Type Ia SNe unreddened in their host galaxies (lowest locus) and for two reddened objects. Based on data discussed by Krisciunas *et al.* (2000).

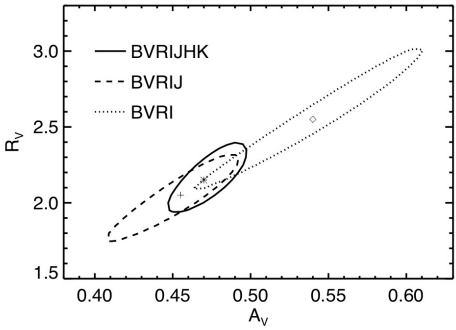

Figure 4. Reddening parameter $R_V$ as a function of the difference of the amounts of *V*-band extinction suffered by SN 2001el and its clone SN 2004S (Krisciunas *et al.* 2007). Figure and analysis method by Peter Garnavich.

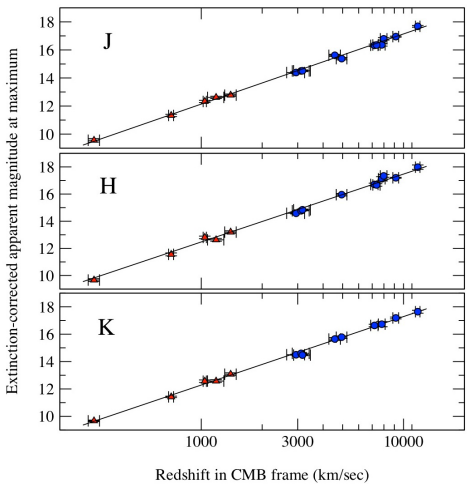

Figure 5. Infrared Hubble diagrams of Type Ia SNe at maximum brightness (Krisciunas *et al.* 2004a). If these objects are standard candles, then from Equation 1 it follows that the y-axis values are a simple function of the logarithm of the distance in parsecs. The slope of the line is fixed at 5.

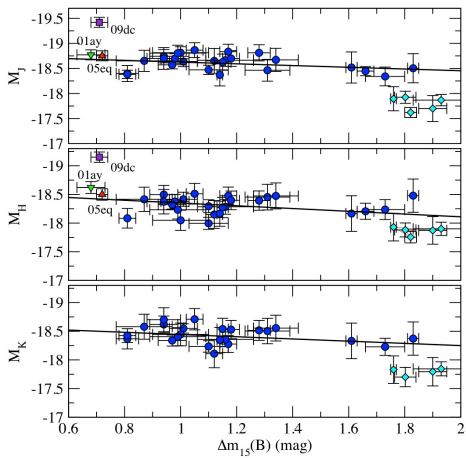

Figure 6. Near-IR absolute magnitudes of Type Ia SNe at maximum brightness, as a function of the *B*-band decline rate parameter (Krisciunas *et al.* 2011).

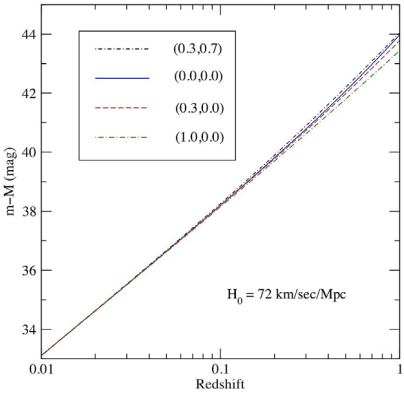

Figure 7. Theoretical Hubble diagram for various combinations of the mass density parameter $\Omega_M$ and the Dark Energy density $\Omega_\Lambda$. The modern "concordance model" of the universe has $\Omega_M \approx 0.3$, $\Omega_\Lambda \approx 0.7$ and a Hubble constant ~ 72 km s$^{-1}$ Mpc$^{-1}$ (Freedman *et al.* 2001). Note that one only starts to detect evidence of Dark Energy from SN photometry at redshift greater than about $z \sim 0.2$.

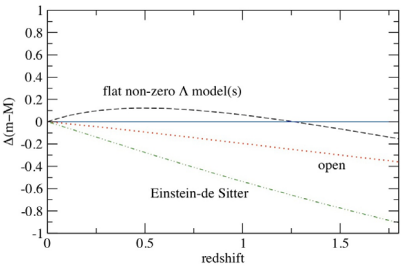

Figure 8. A differential Hubble diagram. We take the "empty universe model" as the reference from Figure 7. Prior to 1998 the expectation was that the data would lie along the "open" line: $\Omega_M = 0.3$, $\Omega_\Lambda = 0.0$. Data of Riess *et al.* (1998) and Perlmutter *et al.* (1999) showed that at redshift ~ 0.5 Type Ia SNe were "too faint" by about 0.2 mag compared to the open model.

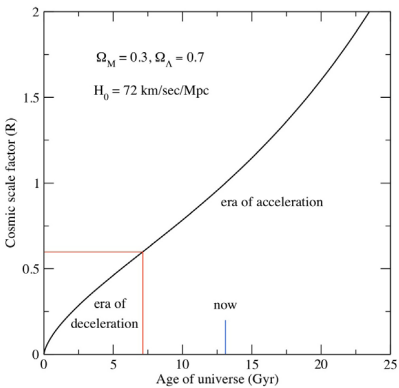

Figure 9. The expansion history of the universe. The y-axis is the cosmic scale factor, effectively the average distance between galaxies.